\documentclass{emulateapj}

\slugcomment{submitted to ApJL}

\shorttitle{Weibel instability with strong magnetic fields}
\shortauthors{Nishikawa et al.}

\begin{document}

\title{
Weibel instability and associated strong fields \\ in a fully 3D  
simulation of a relativistic shock}

\author{K.-I. Nishikawa\altaffilmark{1}, J. Niemiec\altaffilmark{2},
P.E. Hardee\altaffilmark{3}, M. Medvedev\altaffilmark{4}, \\ H. Sol\altaffilmark{5},
Y. Mizuno\altaffilmark{1},
B. Zhang\altaffilmark{6}, M. Pohl\altaffilmark{7},
M. Oka\altaffilmark{1}, D. H. Hartmann\altaffilmark{8}}
  
\altaffiltext{1}{Center for Space Plasma and Aeronomic Research,
University of Alabama in Huntsville, NSSTC, 320 Sparkman Drive,
Huntsville, AL 35805; ken-ichi.nishikawa-1@nasa.gov}
 
\altaffiltext{2}{Institute of Nuclear Physics PAN, ul. Radzikowskiego
152, 31-342 Krak\'{o}w, Poland}

\altaffiltext{3}{Department of Physics and Astronomy, The University
of Alabama, Tuscaloosa, AL 35487}

\altaffiltext{4}{Department of Physics and Astronomy, University of
Kansas, KS 66045}
 
\altaffiltext{5}{LUTH, Observatore de Paris-Meudon, 5 place Jules
Jansen, 92195 Meudon Cedex, France}
  
\altaffiltext{6}{Department of Physics, University of Nevada, Las
Vegas, NV 89154}
    
\altaffiltext{7}{Department of Physics and Astronomy, Iowa State
University, Ames, IA 50011}
    
\altaffiltext{8}{Department of Physics and Astronomy, Clemson
University, Clemson, SC 29634}

\begin{abstract}

Plasma instabilities (e.g., Buneman, Weibel and other two-stream
instabilities) excited in collisionless shocks are responsible for
particle (electron, positron, and ion) acceleration.  Using a new 3-D
relativistic particle-in-cell code, we have investigated the particle
acceleration and shock structure associated with an unmagnetized
relativistic electron-positron jet propagating into an unmagnetized
electron-positron plasma. The simulation has been performed using a
long simulation system in order to study the nonlinear stages of the
Weibel instability, the particle acceleration mechanism, and the shock
structure. Cold jet electrons are thermalized and slowed while the
ambient electrons are swept up to create a partially developed
hydrodynamic (HD) like shock structure. In the leading shock, electron
density increases by a factor of $\lesssim 3.5$ in the simulation
frame. Strong electromagnetic fields are generated in the trailing
shock and provide an emission site. We discuss the possible
implication of our simulation results within the AGN and GRB context.

\end{abstract}

\keywords{relativistic jets: Weibel instability - shock formation -
electron-positron plasma, particle acceleration, magnetic field
generation - particle-in-cell}

\section{Introduction}

Particle-in-cell (PIC) simulations can shed light on the microphysics
within relativistic shocks.  Recent PIC simulations show that particle
acceleration occurs within the downstream jet (e.g., Frederiksen et
al.\ 2004; Nishikawa et al.\ 2003, 2005, 2006, 2008, 2009; Hededal et
al. 2004; Hededal \& Nishikawa 2005; Silva et al.\ 2003; Jaroschek et
al.\ 2005; Chang, Spitkovsky \& Arons 2008; Dieckmann, Shukla, \&
Drury 2008; Spitkovsky 2008a,b; Martins et al.\ 2009).  In general,
these simulations confirm that a relativistic shock in weakly or non
magnetized plasma is dominated by the Weibel instability (Weibel
1959). The associated current filaments and magnetic fields (e.g.,
Medvedev \& Loeb 1999) accelerate electrons (e.g., Nishikawa et
al. 2006) and cosmic rays, which affect the pre-shock medium (Medvedev
\& Zakutnyaya 2009).

In this paper we present new three-dimensional simulation results for
an electron-positron jet injected into an electron-positron plasma
using a long simulation grid. A leading and trailing shock system
develops with strong electromagnetic fields accompanying the trailing
shock.
 
\section{Simulation Setup}

The code used in this study is an MPI-based parallel version of the
relativistic electromagnetic particle (REMP) code TRISTAN (Buneman
1993; Nishikawa et al.\ 2003, Niemiec et al.\ 2008).  The simulations
have been performed using a grid with ($L_{\rm x}, L_{\rm y}, L_{\rm
z}) = (4005, 131, 131)$ cells and a total of $\sim 1$ billion
particles (12 particles$/$cell$/$species for the ambient plasma) in
the active grid.  The electron skin depth, $\lambda_{\rm s} =
c/\omega_{\rm pe} = 10.0\Delta$, where $\omega_{\rm pe} =
(e^{2}n_{\rm a}/\epsilon_0 m_{\rm e})^{1/2}$ is the electron
plasma frequency and the electron Debye length $\lambda_{\rm D}$ is
half of the cell size, $\Delta$. This computational domain is six
times longer than in our previous simulations (Nishikawa et al.\ 2006;
Ramirez-Ruiz, Nishikawa \& Hededal 2007).  The jet-electron number
density in the simulation reference frame is $0.676~n_{\rm a}$, where
$n_{\rm a}$ is the ambient electron density, and the jet Lorentz factor  is
$\gamma_j = 15$.  The jet-electron/positron thermal velocity is
$v_{\rm j,th} = 0.014~c$ in the jet reference frame, where $c=1
$ is the speed of light.  The electron/positron thermal velocity in
the ambient plasma is $v_{\rm a,th} = 0.05~c$. As in our previous work
(e.g., Nishikawa et al.\ 2006) the jet is injected in a plane across
the computational grid located at $x = 25\Delta$ in order to eliminate
effects associated with the boundary at $x = x_{\min}$. Radiating
boundary conditions are used on the planes at  $x =
x_{\min}$ and $x=x_{\max}$ and periodic boundary conditions on all
transverse boundaries (Buneman 1993).

The jet makes contact with the ambient plasma at a 2D interface
spanning the computational domain. Here the formation and dynamics of
a small portion of a much larger shock are studied in a spatial and
temporal way that includes the spatial development of nonlinear
saturation and dissipation from the injection point to the jet front
defined by the fastest moving jet particles.

\section{Simulation Results}

Figure 1a \& b show the averaged (in the $y-z$ plane) (a) jet (red),
ambient (blue), and total (black) electron density and (b)
electromagnetic field energy divided by the total jet kinetic energy
($E^{\rm j}_{\rm t}=\sum _{i = e, p} m_{\rm i}c^{2}(\gamma_{\rm j}
-1)$) at $t = 3250~\omega_{\rm pe}^{-1}$.  Here, ``e'' and ``p'' denote
electron and positron.  Positron density profiles are similar to
electron profiles.
\begin{figure}[h]
%\begin{minipage}[t]{90mm}
\epsscale{0.9}
%\hspace*{0.5cm}
\plotone{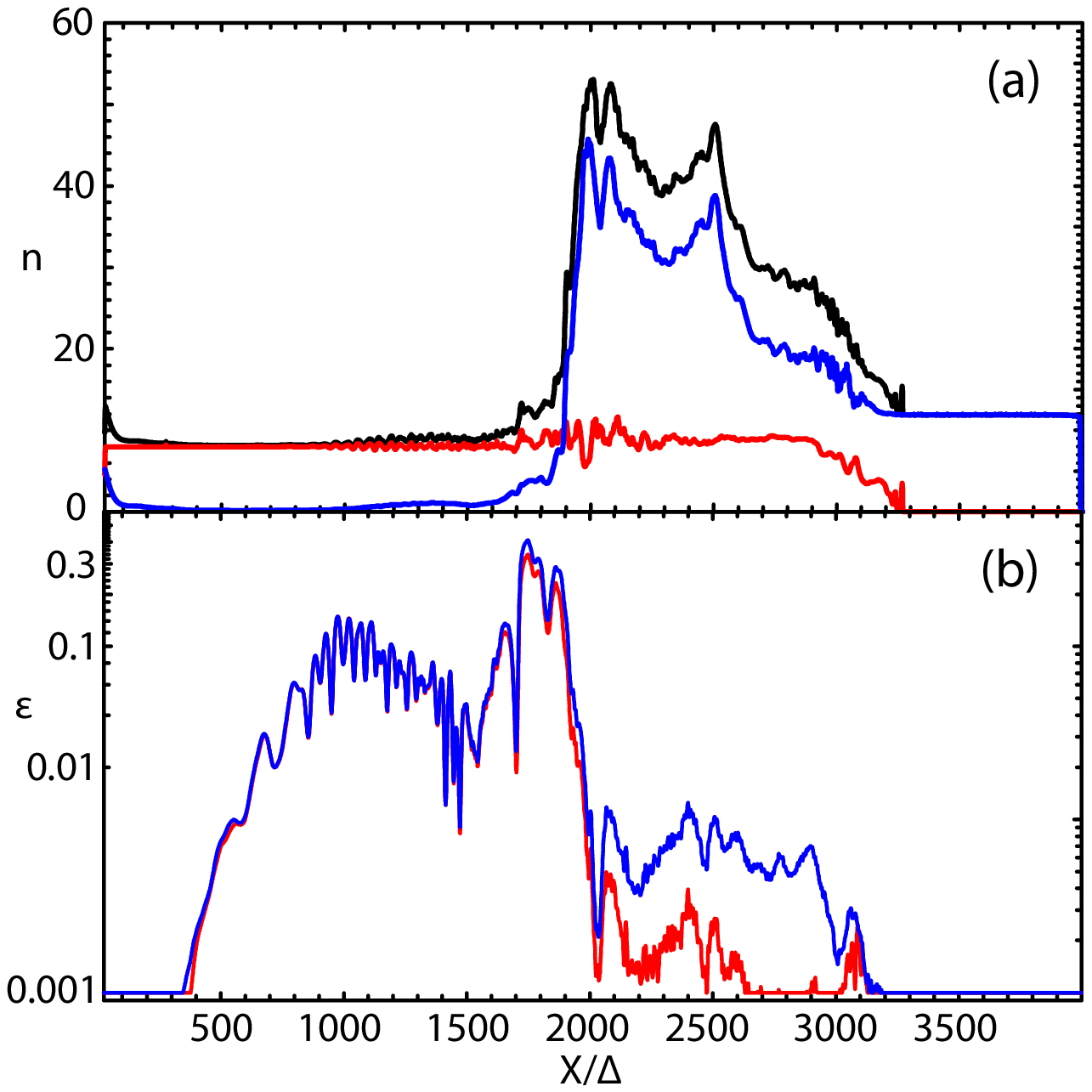}
\plotone{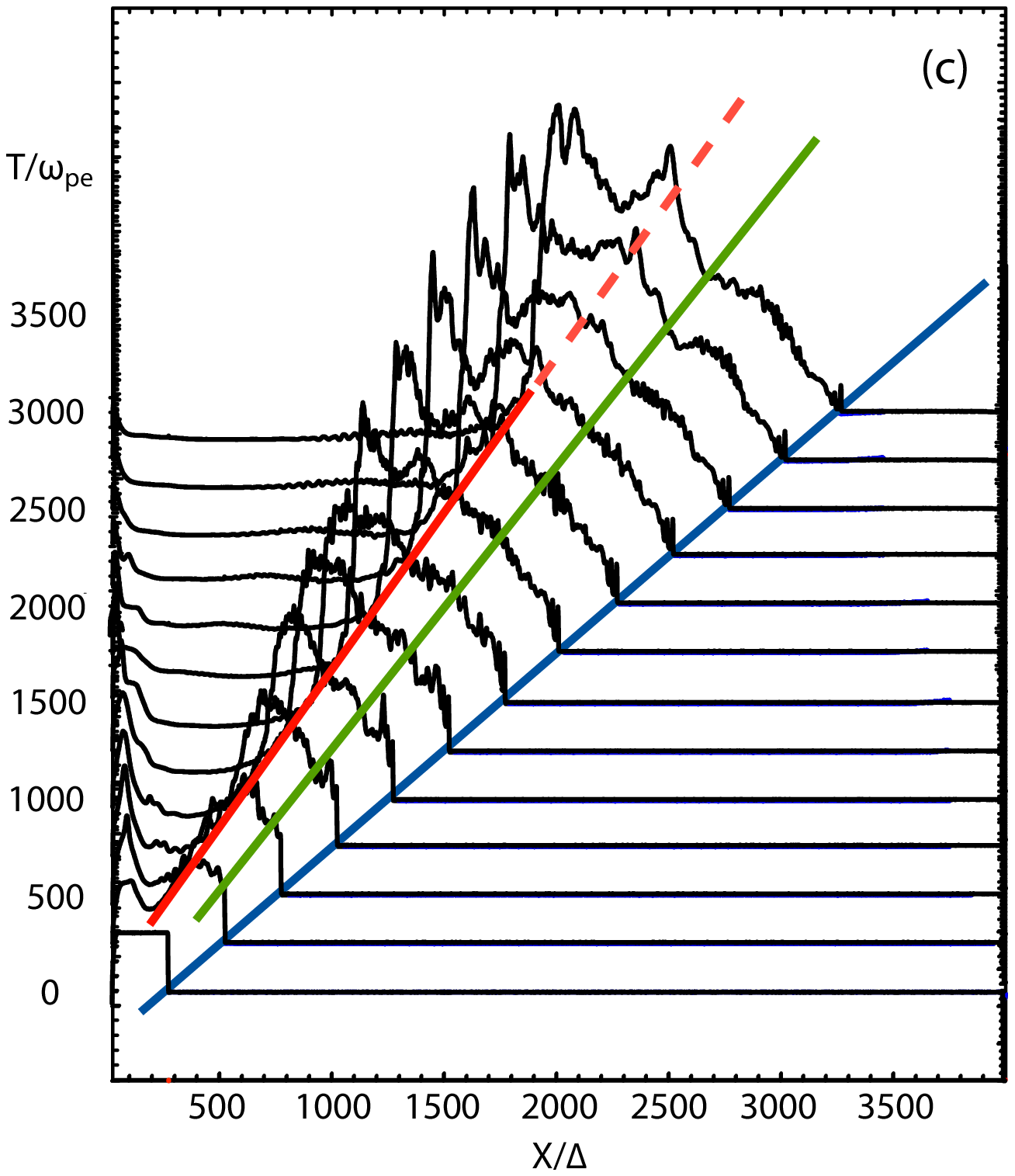}
%\plotone{ded1dATJn05065a.eps}

%\plotone{d1depsB0565a.eps}
%\plotone{Fig1qN.eps}
%\includegraphics[width=180pt, height=110pt]{fig1.eps}
%\end{minipage}
%\begin{minipage}[t]{70mm}
%\vspace*{-6.0cm}
\caption{Averaged values of (a): jet (red), ambient (blue), and total
(black) electron density, and (b): electric (red) and magnetic (blue)
field energy divided by the jet kinetic energy at $t = 3250~\omega_{\rm
pe}^{-1}$.  Panel (c) shows the evolution of the total electron
density in time intervals of $\delta t = 250~\omega_{\rm pe}^{-1}$.  Diagonal
lines indicate motion of the jet front (blue: $\lesssim c$), predicted
contact discontinuity speed (green: $\sim 0.76~c$), and trailing
density jump (red: $\sim 0.56~c$).
\label{fig1}}
%\end{minipage}
\end{figure}
Ambient particles become swept up after jet electrons pass $x/\Delta
\sim 500$.  By $t = 3250~\omega_{\rm pe}^{-1}$, the density has evolved into
a two-step plateau behind the jet front.  The maximum density in this
shocked region is about three times the initial ambient density.  The
jet-particle density remains nearly constant up to near the jet front.

Current filaments and strong electromagnetic fields
accompany growth of the Weibel instability in the trailing shock
region.  The electromagnetic fields are about four times larger than that
seen previously using a much shorter grid system ($L_{\rm x} =
640\Delta$).  At $t = 3250~\omega_{\rm pe}^{-1}$, the electromagnetic
fields are largest at $x/\Delta \sim 1700$, and decline by about one
order of magnitude beyond $x/\Delta = 2300$ in the shocked region
(Nishikawa 2006; Ramirez-Ruiz, Nishikawa \& Hededal 2007).

Figure 1c shows the total electron density plotted at time intervals
of $\delta t = 250~\omega_{\rm pe}^{-1}$. The jet front propagates with the
initial jet speed ($\lesssim c$). Sharp RMHD-simulation shock surfaces
are not created (e.g., Mizuno et al.\ 2009).  A leading shock region
(linear density increase) moves with  a speed between the fastest moving
jet particles $\lesssim c$ and a predicted contact discontinuity speed
of $\sim 0.76~c$ (see \S 4). A contact-discontinuity region consisting
of mixed ambient and jet particles moves at a speed between $\sim
0.76~c$ and the trailing density jump speed $\sim 0.56~c$. A trailing
shock region moves with speed $\lesssim 0.56~c$, note the modest
density increase just behind the large trailing density jump.

\vspace*{-0.0cm}
\begin{figure}[h]
%\begin{minipage}[t]{100mm}
\epsscale{1.2}
\plotone{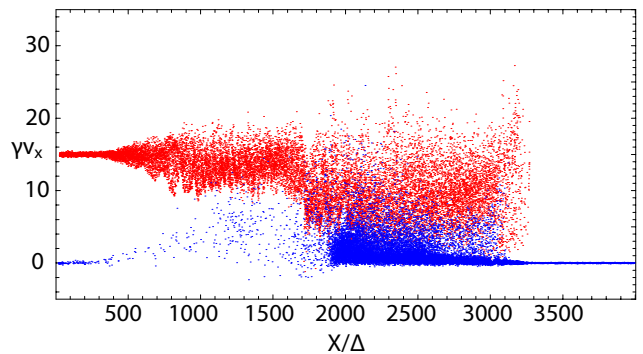}
%\end{minipage}
%\begin{minipage}[t]{60mm}
%\vspace{-5.3cm}
\caption{Phase-space distribution of jet (red) and ambient (blue)
electrons  at $t = 3250~\omega_{\rm
pe}^{-1}$.  About 18,600 electrons of both species are selected
randomly.
\label{fig2}}
% \end{minipage}
\end{figure}

Figure 2 shows the phase-space distribution of jet (red) and ambient
(blue) electrons at $t = 3250~\omega_{\rm pe}^{-1}$ and confirms our
shock-structure interpretation. The electrons injected with $\gamma_j
v_{\rm x} \sim 15$ become thermalized due to Weibel
instabililty-induced interactions.  The swept-up ambient electrons
(blue) are heated by interaction with jet electrons. Some ambient
electrons are strongly accelerated.

%\vspace*{-0.7cm}
\begin{figure}[h]
%\begin{minipage}[t]{100mm}
 \epsscale{1.2}
\plotone{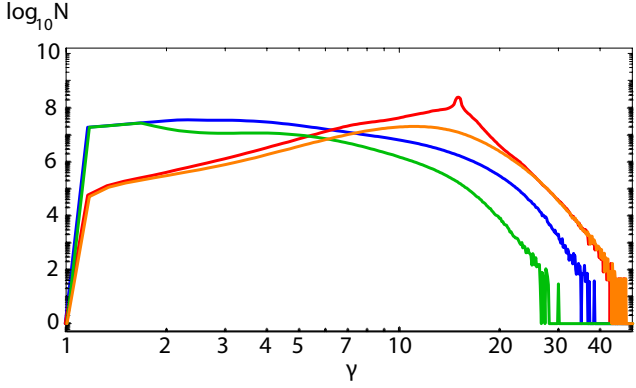}
%\end{minipage}
%\begin{minipage}[t]{60mm}
%\includegraphics[angle=0,scale=.83]{fig3.eps}
%\vspace{-4.8cm}
\caption{Velocity distributions at $t = 3250~\omega_{\rm pe}^{-1}$. All
jet (red) and all ambient (blue), and at $x/\Delta > 2300$ jet
(orange) and ambient (green) electrons are also plotted. The small (red)
peak indicates jet electrons injected at $\gamma_j = 15$.
\label{fig4}}
%\end{minipage}
\end{figure}

Figure 3 shows the velocity distribution of all jet and ambient
electrons in the simulation frame.  The small peak indicates electrons
injected at $\gamma_j = 15$. Jet electrons are accelerated to a
non-thermal distribution. Ambient electrons are also accelerated to
speeds above the jet injection velocity. The velocity distributions of
jet and ambient electrons near the jet front (at $x/\Delta > 2300$) are
also plotted. The fastest jet electrons, $\gamma > 20$, are located near
the jet front. On the other hand, the fastest ambient electrons are
located farther behind the jet front (at $x/\Delta < 2300$). Thus, strong
acceleration of the ambient electrons accompanies the strong fields
associated with the Weibel instability.

\section{Discussion}

Our collisionless-shock structure can be compared to 1-D hydrodynamic
({\bf HD}) shock predictions (e.g., Blandford \& McKee 1976; Zhang \&
Kobayashi 2005).  The speed of the contact discontinuity ({\bf CD}) is
given by ram pressure balance in the CD frame. Our initial conditions
allow us to set the total energy density $e \equiv \rho c^2 +
p/(\Gamma - 1)= \rho c^2$ and pressure $p = 0$, so that the speed in
the ambient frame becomes (Rosen et al.\ 1999)
\begin{equation}
\beta_{\rm cd} = [(\gamma_{\rm j} \eta^{1/2})/(\gamma_{\rm j} \eta^{1/2} +1)] \beta_{\rm j},
\end{equation}
where $\eta \equiv \rho_{\rm j}/\rho_{\rm a} (= m_{\rm e}n_{\rm
j}/m_{\rm e}n_{\rm a})$ and mass densities are determined in the
``jet'' and ``ambient'' proper frames. In the simulation $n_{\rm j} =
0.0451 n_{\rm a}$ and $\gamma_{\rm j} = 15$, and $\beta_{\rm cd} =
0.759$ ($\gamma_{\rm cd} = 1.54$) is the predicted CD speed. Formally
this should represent the average speed of particles in the CD region.

The leading shock moves at a speed given by 
\begin{equation}
\gamma_{\rm ls}^2 =
{{(\gamma_{\rm cd}+1)[\Gamma_{\rm sa}(\gamma_{\rm cd}-1)+1]^2}\over{\Gamma_{\rm sa}(2-\Gamma_{\rm sa})(\gamma_{\rm cd}-1)+2}}
\end{equation}
where $5/3 > \Gamma_{\rm sa} > 4/3$ is the shocked ambient adiabatic
index. Thus the leading shock speed is predicted to be $0.865 >
\beta_{\rm ls} > 0.783$ ($ 2 > \gamma_{\rm ls} > 1.6$) where upper and
lower limits correspond to upper and lower limits of $\Gamma_{\rm sa}$,
respectively.

The jump condition at the leading shock is 
\begin{equation}
{n_{\rm sa} \over n_{\rm a}} = {\Gamma_{\rm sa} \gamma_{\rm cd} + 1 \over \Gamma_{\rm sa} - 1},
\end{equation}
where $n_{\rm sa}$ is the shocked ambient density in the proper (CD) frame
and we find $5.34~n_{\rm a} < n_{\rm sa} < 9.15~n_a$, where the lower and upper
limits correspond to the upper and lower limits to $\Gamma_{\rm sa}$,
respectively.  Measured in the ambient (simulation) frame the shocked
ambient density should be $8.2~n_{\rm a} < \gamma_{\rm cd} n_{\rm sa} <
14.1~n_{\rm a}$. Formally this should represent the total density of
particles in the shocked-ambient region.

Computations associated with the trailing shock are most easily
performed in the jet rest frame designated below as the ``primed''
frame.  In this frame the CD moves with speed
$
\beta'_{\rm cd} = -(\beta_{\rm j} - \beta_{\rm cd})/( 1 - \beta_{\rm j}\beta_{\rm cd} ) = -0.984
$
and $\gamma'_{\rm cd} = 5.60$. The speed of the trailing shock in the
jet frame, $\gamma'_{\rm ts}$ is given by eq.\ (2) but with
$\gamma_{\rm cd} \rightarrow \gamma'_{\rm cd}$ and $\Gamma_{\rm sa}
\rightarrow
\Gamma_{\rm sj}$ where $\Gamma_{\rm sj}$ is the shocked-jet adiabatic 
index. In the jet frame $10.4 > \gamma'_{\rm ts} > 7.4$ and $0.995 >
-\beta'_{\rm ts} > 0.991$, where upper and lower limits correspond to upper 
$\Gamma_{\rm sj} = 5/3$ and lower $\Gamma_{\rm sj} = 4/3$ limits to
$\Gamma_{\rm sj}$, respectively.  The trailing shock speed in the
ambient (simulation) frame is 
$ 0.35 < \beta_{\rm ts} = (\beta_{\rm j} - \beta'_{\rm ts}) / (1 -
\beta_{\rm j}\beta'_{\rm ts}) < 0.61 
$
where the lower and upper limits correspond
to the upper and lower limits of $\Gamma_{\rm sj}$, respectively.

The density jump at the trailing shock is given by eq.\ (3) but with
$\gamma_{\rm cd} \rightarrow \gamma'_{\rm cd}$ and $\Gamma_{\rm sa}
\rightarrow
\Gamma_{\rm sj}$ where now $n_{\rm sa}/n_a \rightarrow n_{\rm sj}/n_{\rm j}$ where 
$n_{\rm j} = 0.0451~n_{\rm a}$ with result that the proper density of
shocked jet material is $0.70~n_a < n_{\rm sj} < 1.15~n_{\rm a}$ where
lower and upper limits correspond to upper and lower limits to
$\Gamma_{\rm sj}$, respectively.  In the ambient (simulation) frame
the shocked jet density should be $1.08~n_{\rm a} < \gamma_{\rm cd}
n_{\rm sj} < 1.76~n_{\rm a}$. Formally this should represent the total
density of particles in the shocked jet region.

In the simulation the speed of the trailing density jump is $\sim
0.56~c$, which is in the predicted range $0.35 < \beta_{\rm ts} <
0.61$, a typical speed within the density-plateau region, $\sim
0.75~c$, is close to $\beta_{\rm cd}= 0.76$. The poorly defined
leading shock structure moves at a speed between $\sim 0.76~c$ and
$\lesssim c$, consistent with the predicted $0.78 <
\beta_{\rm ls} < 0.86$.

In the simulation the maximum density increase observed in the ambient
(simulation) frame is $\gamma_{\rm cd} n_{\rm sa}/n_{\rm a} \sim 3.5$
behind the leading shock (see Fig. 1a).  This is about a factor of
$\sim 3$ smaller than the predicted increase, $8.2 < \gamma_{\rm cd}
n_{\rm sa}/n_{\rm a} < 14.1$, for a fully-developed leading shock. On
the other hand, the density increase observed in the ambient
(simulation) frame of $\gamma_{\rm cd} n_{\rm sj}/n_{\rm a} \gtrsim 1$
just before the trailing large density jump is comparable to that
predicted, $1.08 < \gamma_{\rm cd} n_{\rm sj}/n_{\rm a} < 1.76$, for a
fully developed trailing shock.
 
Our present results can be compared to those found in the 2-D
simulations of Chang et al.\ (2008) (see also Spitkovsky 2008a).
Their simulations were performed in the {\bf
CD} frame, and material with proper density, n, moved into the contact
discontinuity with a Lorentz factor $\gamma = 15$. A shock moved away
from the CD with the predicted speed
\begin{equation}
\beta_{\rm s} = (\Gamma_{\rm s} - 1)\left[{\gamma - 1 \over \gamma +1}\right]^{1/2} = 0.47~,
\end{equation}
and predicted density jump
\begin{equation}
{n_{\rm s} \over \gamma n} = {1 \over \gamma}{\Gamma_{\rm s} \gamma + 1 \over \Gamma_{\rm s} - 1} = 3.13~,
\end{equation}
for a shocked adiabatic index of $\Gamma_{\rm s} = 3/2$.

In our simulation we have two shocks that move away from the CD.  For
our leading shock, the ambient  plasma moves relative to the CD at a speed equal to
$\beta_{\rm cd} = 0.759$ and $\gamma = \gamma_{\rm cd} = 1.54$ in eqs. 4 \&
5. In the CD frame $\beta_{\rm s} = 0.23$ and the observed density jump
becomes $n_{\rm sa}/\gamma_{\rm cd} n_a = 4.3$ for $\Gamma_{\rm s} = 3/2$. So we
see that our leading shock speed would be about 50\% less than that in
Chang et al.\ (2008) and our density increase would be about 50\% larger
for a fully-developed leading shock in the CD frame.  For the trailing
shock, the jet moves  toward the CD at a speed equal to $-\beta'_{\rm cd} =
0.984$ and $\gamma =\gamma'_{\rm cd} = 5.60$ in eqs. 4 \& 5.  In the CD
frame $\beta_{\rm s} = 0.417$ and the observed density increase becomes
$n_{sj}/\gamma'_{\rm cd} n_j = 3.36$ for $\Gamma_{\rm s} = 3/2$.  So we see that
our trailing shock speed would be about 11\% less than that in Chang
et al.\ (2008) and our density increase would be about 7\% larger for the
fully developed trailing shock in the CD frame.  The parameters
associated with our trailing shock are similar to those found in Chang
et al.\ (2008), and  the Weibel filamentation structures are comparable but now studied  in 
full 3-D.

\section{Conclusion}

The present simulation finds for the first time a relativistic shock
system comparable to a predicted relativistic HD shock system
consisting of leading and trailing shocks separated by a contact
discontinuity, albeit not yet fully developed. One remarkable aspect
of this shock system lies in the generation of large electromagnetic
fields, up to 30\% of the kinetic energy density, associated with the
trailing shock.  Electromagnetic fields in the leading shock and
contact-discontinuity region are over one order of magnitude lower.
The large value for $\epsilon_B \sim 0.3$ in our trailing shock hints
that Poynting-flux-dominated ejecta may  not be required to explain
some GRB observations (McMahon et al.\ 2006).

Visualization of our dual shock system in the ambient (simulation)
frame provides a picture of the shock structure that should exist at
the head of a relativistic astrophysical jet, $\gamma_{\rm jt} = 15$,
that is less dense than the surrounding medium, $n_{\rm jt}/n_{\rm am}
= 0.045$.  Within the AGN context, here we identify our trailing shock with the ``jet'' shock
that decelerates the relativistic jet and we would expect synchrotron
emission to originate from the strongly magnetized structure.  Little
synchrotron emission would originate from the weakly magnetized
``bow'' shock in front of the contact discontinuity. This in fact is
what is observed at the leading edge of extra-galactic jets where
synchrotron emission from the bow shock is not typically observed.

Visualization of our dual shock system in the ``jet'' frame provides a
picture of the shock structure that would accompany a relativistic
blast wave driven by relativistic ejecta.  Within the GRB context, here we identify the
ambient medium as representing relativistic ejecta moving at
$\gamma_{\rm ej} = 15$ into a much less dense ISM, $n_{\rm ej}/n_{\rm
ism} = 22$. Our trailing shock is now identified with the ``forward''
shock and we would expect synchrotron emission from this strongly
magnetized structure. Little synchrotron emission would originate from
the low Lorentz factor, weakly-magnetized ``reverse'' shock moving
back into the ejecta.

Our present simulation involves an electron-positron jet and ambient
medium.  We might expect similar shock-structure development in
electron-ion simulations, albeit on much longer temporal and spatial
scales.
   
\acknowledgments
This work is supported by AST-0506719, AST-0506666, NASA-NNG05GK73G,
NNX07AJ88G, NNX08AG83G, NNX08AL39G, and NNX09AD16G.  JN is supported
by MNiSW research project N N203 393034, and The Foundation for Polish
Science through the HOMING program, which is supported by a grant from
Iceland, Liechtenstein, and Norway through the EEA Financial
Mechanism.  Simulations were performed at the Columbia facility at the
NASA Advanced Supercomputing (NAS) and Cobalt at the National Center
for Supercomputing Applications (NCSA) which is supported by the NSF.
Part of this work was done while K.-I. N. was visiting The
Observatoire de Paris, Meudon in summer of 2008. Support from the
French Natural Science Research Council is gratefully acknowledged.

\end{document}